%% file: main.tex
\documentclass[sigconf]{acmart}

\AtBeginDocument{%
  }

\setcopyright{acmcopyright}
\copyrightyear{2024}
\acmYear{2024}
\acmDOI{XXXXXXX.XXXXXXX}
\acmPrice{15.00}
\acmISBN{978-1-4503-XXXX-X/18/06}
\usepackage{listings}
\input{macros}

\begin{document}

\definecolor{codegreen}{rgb}{0,0.6,0}
\definecolor{codegray}{rgb}{0.5,0.5,0.5}
\definecolor{codepurple}{rgb}{0.58,0,0.82}
\definecolor{backcolour}{rgb}{0.95,0.95,0.92}

\definecolor{javakeyword}{RGB}{0,0,255}
\definecolor{javastring}{RGB}{225,0,0}
\definecolor{javacomment}{RGB}{0,128,0}
\definecolor{javaannotation}{RGB}{102,153,0}

\lstdefinestyle{mystyle}{
    language=Java,
    aboveskip=3mm,
    showstringspaces=false,
    columns=flexible,
    backgroundcolor=\color{backcolour},
    keywordstyle=\color{javakeyword}\bfseries,
    stringstyle=\color{javastring},
    commentstyle=\color{javacomment}\itshape,
    basicstyle=\ttfamily\small,
    numberstyle=\tiny\color{codegray},
    breakatwhitespace=false,         
    breaklines=true,                 
    captionpos=b,                    
    keepspaces=false,   
    numbers=left,              
    numbersep=5pt,                  
    showspaces=false,                
    showstringspaces=false,
    showtabs=false,                  
    tabsize=2,
    escapeinside=``,
    literate=
        {@Override}{{{\color{javaannotation}@Override}}}1
        {@Test}{{{\color{javaannotation}@Test}}}1
}


\lstset{style=mystyle}

\title[]{\textsc{StaAgent}: An Agentic Framework for Testing Static Analyzers}

\author{Elijah Nnorom}
\email{ennorom@yorku.ca}
\affiliation{%
	\institution{York University}
	\streetaddress{4700 Keele Street}
	\city{Toronto}
	\state{ON}
	\country{Canada}
	\postcode{M3J 1P3}
}

\author{Md Basim Uddin Ahmed}
\email{mbahmed@yorku.ca}
\affiliation{%
  \institution{York University}
  \streetaddress{4700 Keele Street}
  \city{Toronto}
  \state{ON}
  \country{Canada}
  \postcode{M3J 1P3}
}

\author{Jiho Shin}
\email{jihoshin@yorku.ca}
\affiliation{%
	\institution{York University}
	\streetaddress{4700 Keele Street}
	\city{Toronto}
	\state{ON}
	\country{Canada}
	\postcode{M3J 1P3}
}

\author{Hung Viet Pham}
\email{hvpham@yorku.ca}
\affiliation{%
	\institution{York University}
	\streetaddress{4700 Keele Street}
	\city{Toronto}
	\state{ON}
	\country{Canada}
	\postcode{M3J 1P3}
}

\author{Song Wang}
\email{wangsong@yorku.ca}
\affiliation{%
	\institution{York University}
	\streetaddress{4700 Keele Street}
	\city{Toronto}
	\state{ON}
	\country{Canada}
	\postcode{M3J 1P3}
}

\input{sections/abstract}



\received{28 September 2024}
\received[revised]{5 March 2025}
\received[accepted]{16 April 2025}

\maketitle

\input{sections/intro}
\input{sections/background}
\input{sections/approach}
\input{sections/experiment}

\input{sections/results}

\input{sections/threats}
\input{sections/conclusion}

\bibliographystyle{ACM-Reference-Format}

\bibliography{paper}



\end{document}

%% file: macros.tex
\usepackage{soul}
\usepackage{multirow}
\usepackage{listings}
\usepackage{fancybox}
\usepackage{enumitem}
\usepackage{graphicx}
\usepackage{color}
\usepackage{xcolor}
\usepackage{array}
\usepackage{amsmath}
\usepackage{xspace}
\usepackage{url}
\usepackage{tikz}
\usepackage{caption}
\usepackage{pgfplots}
\usepackage{balance}
\usepackage{subcaption}
\pgfplotsset{compat=1.16}
\usepackage{pgf-pie}
\usetikzlibrary{pgfplots.statistics,calc}
\usepackage{algorithm}
\usepackage{algorithmic}
\usepackage{adjustbox}
\usepackage{makecell}
\usepackage{comment}

\usepackage{tcolorbox}
\makeatletter
\newcommand{\mybox}[1]{%
	\setbox0=\hbox{#1}%
	\setlength{\@tempdima}{\dimexpr\wd0+13pt}%
	\begin{tcolorbox}[boxrule=0.5pt, colback=white, arc=4pt,
		left=6pt,right=6pt,top=6pt,bottom=6pt,boxsep=0pt]
		#1
	\end{tcolorbox}
}

\newcommand{\tool}{\textsc{StaAgent}}

%% file: sections/abstract.tex
\begin{abstract}

Static analyzers play a critical role in identifying bugs early in the software development lifecycle, but their rule implementations are often under-tested and prone to inconsistencies. To address this, we propose \textsc{StaAgent}, an agentic framework that harnesses the generative capabilities of Large Language Models (LLMs) to evaluate static analyzer rules systematically. \textsc{StaAgent} comprises four specialized agents: a \textit{Seed Generation Agent} that translates bug detection rules into concrete, bug-inducing seed programs; a \textit{Code Validation Agent} that ensures the correctness of these seeds; a \textit{Mutation Generation Agent} that produces semantically equivalent mutants; and an \textit{Analyzer Evaluation Agent} that performs metamorphic testing by comparing the static analyzer’s behavior on seeds and their corresponding mutants. By revealing inconsistent behaviors, \textsc{StaAgent} helps uncover flaws in rule implementations. This LLM-driven, multi-agent framework offers a scalable and adaptable solution to improve the reliability of static analyzers.

We evaluated \textsc{StaAgent} with five state-of-the-art LLMs (CodeLlama, DeepSeek, Codestral, Qwen, and GPT-4o) across five widely used static analyzers (SpotBugs, SonarQube, ErrorProne, Infer, and PMD).
The experimental results show that our approach can help reveal 64 problematic rules in the latest versions of these five static analyzers (i.e., 28 in SpotBugs, 18 in SonarQube, 6 in ErrorProne, 4 in Infer, and 8 in PMD). In addition, 53 out of the 64 bugs (i.e., 83\%) cannot be detected by the state-of-the-art baseline. We have reported all the bugs to developers, with two of them already fixed. Three more have been confirmed by developers, while the rest are awaiting response. 
These results demonstrate the effectiveness of our approach and underscore the promise of agentic, LLM-driven data synthesis to advance software engineering. 
\end{abstract}

%% file: sections/intro.tex
\section{Introduction}
\label{sec:intro}
Static analyzers are now integral to modern software development to automatically detect potential bugs, vulnerabilities, and coding issues early in the development lifecycle with well-defined bug rules~\cite{nachtigall2022large, li2024enhancing, wang2022find}.
Their ability to analyze source code without execution makes them widely adopted in industry. 
However, despite their widespread use, static analyzers often exhibit inconsistent behaviors, false negatives, and blind spots, leading to a lack of trust and reduced utility in real-world applications. These limitations are exacerbated by the complexity of modern programming languages and the diverse coding practices used by developers.

A key challenge in improving static analyzers lies in effectively evaluating their correctness and reliability. Traditional validation methods, such as manual rule testing~\cite{croft2021empirical,sun2012extending} and curated bug datasets~\cite{mordahl2023automatic, nunes2018benchmarking}, are often insufficient, as they fail to cover the wide spectrum of possible code variations.
Metamorphic testing has emerged as a promising approach to address this problem by identifying inconsistencies in an analyzer when analyzing inputs with equivalent semantics~\cite{wang2022find}. 
Recently, Zhang et al.~\cite{zhang2023statfier} proposed Statfier, which employs metamorphic testing by reusing human-created seed programs from tool test suites and documentation, then applying semantic-preserving transformations to generate mutants and detect inconsistencies between seeds and mutants. 
However, a key limitation of this approach is the challenge of generating diverse and semantically equivalent code snippets. These examples often require manual curation or extraction from existing test suites, making the process labor-intensive. Furthermore, such strategies tend to lack generalizability, as they are closely tied to specific tools, languages, or domains.  

To bridge this gap, we propose an agentic framework, \textsc{StaAgent}, comprising four specialized agents: a \textit{Seed Generation Agent}, which synthesizes bug-inducing code snippets; a \textit{Code Validation Agent}, which ensures the correctness of seed bugs by generating tests that verify the behavior aligns with the rule description; a \textit{Mutation Generation Agent}, which applies semantic-preserving transformations to produce mutants; and an \textit{Analyzer Evaluation Agent}, which automatically assesses the correctness of static analyzer rules through metamorphic testing by comparing the analyzer’s behavior on seeds and their corresponding mutants. 
\textsc{StaAgent} is designed for scalability, allowing new rules and mutation operators to be incorporated with minimal manual effort. Furthermore, leveraging the broad language and domain knowledge embedded in modern LLMs, the framework is adaptable to diverse static analyzers and programming ecosystems. 

We evaluate \textsc{StaAgent} with five state-of-the-art LLMs (CodeLlama, DeepSeek, Codestral, Qwen, and GPT-4o) across five widely used static analyzers (SpotBugs, SonarQube, ErrorProne, Infer, and PMD). The experimental results demonstrate that \textsc{StaAgent} can help reveal 64 problematic rules in the five analyzers (i.e., 28 in SpotBugs, 18 in SonarQube, 6 in ErrorProne, 4 in Infer, and 8 in PMD). We have reported all the identified bugs to developers, with two of them already fixed. Three more have been confirmed by developers, while the rest are awaiting response. 
These results demonstrate the effectiveness of our approach and underscore the promise of agentic, LLM-driven data synthesis for advancing software engineering. 

This paper makes the following contributions:
\begin{itemize}[leftmargin=*]
    \item We propose \textsc{StaAgent}, an innovative agentic testing framework that leverages LLMs to automatically generate, mutate, and validate code snippets for testing static analyzers. 
    \item We present a comprehensive study involving five LLMs and five static analyzers, which led to the discovery of 64 faulty rule implementations in total, all of which have been reported to the respective developers.
    \item  We demonstrate the effectiveness of agentic LLM-driven data synthesis in uncovering inconsistencies in static analyzer rules, highlighting its potential to advance software testing tasks. 
    \item We release the data and source code of our experiments to enable other researchers to replicate and extend our study\footnote{\url{https://anonymous.4open.science/r/LLM-For-Testing-Static-Tools-D1BF/}}.
\end{itemize}


%% file: sections/background.tex
\section{Background and Related Work}
\label{sec:background}

\subsection{LLM-based Data Synthesis}

The advent of deep learning has amplified the demand for high‐quality data, especially in data‐scarce domains~\cite{whitehouse2023llm}. Large language models now underpin diverse data‐curation tasks—ranging from health‐analytics synthesis~\cite{long2024llms} and cross‑lingual augmentation~\cite{whitehouse2023llm} to low‑resource language augmentation~\cite{joel2024survey}, tabular data generation~\cite{wang2025harmonic}, and automated benchmark creation~\cite{shao2025case2code}. Numerous studies report that LLM‐generated text and code can rival human quality~\cite{long2024llms,liu2023your} and yield greater variety~\cite{chung2023increasing}. In software engineering, LLMs have driven advances in test‑driven development~\cite{mathews2024test,fakhoury2024llm}, synthetic benchmark synthesis~\cite{shao2025case2code}, and compiler testing-Gu’s CodeT5-based Golang tests achieved 3.38\% coverage versus GoFuzz’s 0.44\%\cite{gu2023llm}. Feedback‑driven frameworks further refine generation accuracy, with user-in-the-loop methods yielding up to 46\% gains\cite{fakhoury2024llm} and ClarifyGPT~\cite{mu2024clarifygpt} improving MBPP performance by 15\% on the MBPP dataset~\cite{austin2021program}. More recently, PlanSearch has outperformed standard sampling by leveraging high‑level planning for code synthesis~\cite{wang2024planning}.

Our work utilizes LLMs and the feedback mechanism to better generate semantically equivalent code to synthesize data, which enables metamorphic testing of static analyzer rules with minimum human intervention.

\subsection{Static Analysis}

Static analysis identifies software bugs and vulnerabilities by examining code properties without execution~\cite{rival2020introduction}. While these tools use predefined rules to find code deviations, they frequently generate many false positives and false negatives~\cite{thomas2016questions}, making rigorous evaluation essential. This section will discuss current methods for detecting static analyzer flaws and contrast them with our approach.

A recent and closely related work by \emph{Zhang et al.} employs metamorphic testing by harvesting seed programs from existing test suites and documentation, then applying semantics‑preserving transformations~\cite{zhang2023statfier}.
Our method instead uses LLMs to directly generate seeds from analyzer rule descriptions, reducing manual effort and removing dependence on pre-existing test suites. This approach also simplifies adding new mutation operators by eliminating the need for individual mutation implementations.
Fuzz4All also uses LLMs but targets diverse execution paths and runtime anomalies rather than rule‑specific faults, so it lacks the fine‑grained localization and complete rule coverage our approach provides~\cite{xia2024fuzz4all}. Other LLM-based approaches include UBITect \cite{li2024enhancing}, which improves tool precision by purging execution paths, though its generalizability beyond UBI-related bugs is limited. Additionally, Mohajer et al. \cite{mohajer2024effectiveness} showed ChatGPT's effectiveness in identifying static analyzer false positives. 
Other approaches have also employed various automated and semi-automated techniques. Fuzz‑based and oracle‑driven methods such as Csmith‑style random generators~\cite{cuoq2012testing}, constraint‑based oracles~\cite{fleischmann2024constraint}, interrogation testing~\cite{kaindlstorfer2024interrogation}, and automated oracle construction~\cite{he2024finding} produce large suites of C/C\texttt{++} programs to expose soundness or precision flaws. However, they cannot localize defects to specific rules. Q-SAST follows a similar strategy but requires 540 person‑hours of manual inspection and relies on C-specific mutations that do not generalize to Java~\cite{li2024evaluating}. $\alpha$-Diff identifies unsoundness by cross-tool comparison rather than analyzing faults within a single tool~\cite{klinger2019differentially}.
Domain‑specific bug‑injection frameworks for IoT~\cite{parveen2020mutation} and smart contracts~\cite{ghaleb2020effective} insert hand-crafted vulnerabilities into benign code to gauge precision and recall. However, this demands expert implementation, is costly to scale, limits mutation diversity, and restricts applicability to target domains.
Some contemporary works, such as SAScope~\cite{zhang2024characterizing}, annotation-induced fault detection~\cite{zhang2024understanding}, and testing compiler~\cite{taneja2020testing} or kernel~\cite{vishwanathan2022sound} optimizers, are orthogonal to our research. 
Similarly, \emph{Zhang et al.}~\cite{zhang2024understanding} only addresses annotation-related issue detectors, which is complementary to our work.


In summary, our approach provides scalability in terms of rule and tool coverage, effortless extensibility for mutation operators, and generalizability due to its independence from language or domain-specific constructs. Unlike existing methods, it can also provide granular fault localization, significantly reducing manual triaging efforts associated with identified defects.

%% file: sections/approach.tex
\section{Approach}
\label{sec:emp_data}
\input{fig/overview}

Figure~\ref{fig:overview} depicts the workflow of \textsc{StaAgent}. Starting with a given static analyzer, we first extract its bug detection rules from its online documentation (Section~\ref{sec:3.1}). These rules are then passed to the \textit{Seed Generation Agent}, which generates bug-inducing seed programs (Section~\ref{sec:3.2}). Next, the \textit{Code Validation Agent} leverages an LLM to create tests that trigger the seed bugs and verifies whether the detected bugs align with their respective rule descriptions, based on both the seed code and test outcomes (Section~\ref{sec:3.3}). The \textit{Mutation Generation Agent} then uses the LLM to produce semantically equivalent mutants of the seeds through a set of mutation operators (Section~\ref{sec:3.4}), which undergo similar testing and validation. Finally, the \textit{Analyzer Evaluation Agent} runs the static analyzer on both the seed bugs and their mutants (Section~\ref{sec:3.5}), flagging any inconsistent behavior between them as potential flaws in the analyzer’s rule implementations. 

\subsection{Bug Rule Collection}
\label{sec:3.1}
The core idea of this study is to detect bugs in static analyzers through agentic, LLM-driven metamorphic testing, which involves providing semantically equivalent inputs to the analyzer and identifying inconsistencies in its outputs. To achieve this, we treat each bug detection rule implemented in the static analyzer as a distinct functional component of the tool. 
To collect the rules of a static analyzer, we developed automated crawlers to collect and parse the bug detection rules from the analyzer’s official documentation, online resources, and rule repositories. These crawlers systematically extract information such as rule names, rule descriptions, example violations, and severity levels. 
Our study focuses exclusively on bug detection rules related to code correctness, security vulnerabilities, and performance, as these directly impact the reliability and robustness of software systems. We intentionally exclude rules related to best practices, coding style, and maintainability since such rules are often subjective, do not directly indicate functional issues, and may vary significantly between different static analyzers. We also exclude rules that only apply to specific third-party libraries, such as Android APIs and the Spring Framework, as these rules are framework-specific and may not generalize to broader development tasks. Finally, we omit rules that target test code, as they primarily focus on test structure, assertions, and mocking practices, rather than detecting functional defects in production code.


\input{fig/spotbugs_19}

Figure~\ref{fig:ruleExamp} presents an example rule extracted from SpotBugs' online rule documentation. This rule detects potential overflow issues when computing the absolute value of \texttt{hashCode}. Specifically, if \texttt{hashCode} returns \texttt{Integer.MIN\_VALUE} (i.e., -2147483648), applying the absolute value function results in an overflow, as 2147483648 exceeds the maximum representable 32-bit integer (2147483647). Consequently, SpotBugs flags this operation as a bug when the absolute value method is applied to \texttt{hashCode}.



\subsection{Seed Generation Agent}
\label{sec:3.2}

Given a bug detection rule from a static analyzer, we input the rule title and description into this agent to generate a corresponding seed bug program that contains the described bug. This ensures that the generated code accurately reflects the intended rule violation. To enhance usability and maintainability, the generated bug-containing program must meet the following criteria:
\begin{itemize}[leftmargin=*]
    \item \textbf{Compilability}: The seed bug program must be syntactically correct and able to compile without errors, ensuring that the static analyzer can process it effectively.
    \item \textbf{Self-Containment}: The bug should be isolated within a single function (e.g., \texttt{showBug()}), avoiding unnecessary dependencies and keeping the program minimal by omitting complex logic or extraneous code.
    \item \textbf{No Third-Party Libraries}: The code should only use standard Java libraries, as third-party dependencies increase maintenance overhead and may introduce unintended interactions that affect analysis.
\end{itemize}
By following these constraints, we aim to generate seed bug programs that are effective, reproducible, and easy to maintain. The detailed prompt templates are available in our replication package.  


Note that LLMs can generate code with syntax errors for various reasons, including the use of outdated Java APIs, lack of strong syntax enforcement, incorrect imports or library usage, and hallucinated or non-existent methods~\cite{chen2024deep, tambon2025bugs}. 
For each generated seed bug program, this agent also performs a compilation check. If the code fails to compile, the code and the error message will be fed back into the LLM, prompting it to revise the code to resolve the compilation issue. We repeat this iterative refinement process up to five times. If the code remains uncompilable after five attempts, we discard the candidate seed bug and generate a new one. This ensures that only valid, compilable bug instances are retained for further analysis.  
We maintain a metadata file for each seed bug program that records key information about the bug, including the corresponding bug detection rule and the specific buggy lines in the code. 
Listing \ref{code:seed} shows an example of a seed bug generated by the agent (with GPT-4o) to demonstrate a faulty hashCode calculation shown in Figure~\ref{fig:ruleExamp}.

\begin{lstlisting}[style=mystyle, label={code:seed}, caption={A seed program generated by the \textit{Seed Generation Agent} to demonstrate the problem of calculating absolute value of hashCode, as shown in Figure \ref{fig:ruleExamp}.}]
public class HashCodeBug {
    public int showBug(String input) {
        int hashcode = input.hashCode();
        int absHashcode = Math.abs(hashcode);
        return absHashcode;
    }
}
\end{lstlisting}

\subsection{Code Validation Agent}
\label{sec:3.3}
To verify the presence of a potential bug in the seed program, this agent uses an LLM to generate test cases designed to trigger the bug. Similar to the seed bug generation process, test cases generated by LLMs may also contain syntax errors or other inconsistencies. 
Thus, we perform a compilation check for each generated test case. If a test case fails to compile with its corresponding seed buggy program, we feed the test case, seed program, and error traceback into the LLM, prompting it to revise the test case to resolve the compilation issue. This iterative refinement process continues for a predefined number of attempts (e.g., five iterations). If the test case remains uncompilable after multiple revisions, it is discarded, and a new test case is generated. Listing \ref{code:test} shows the test case generated for the seed program shown in Listing \ref{code:seed}.

We further use the test case to verify the alignment between a seed bug and its corresponding bug detection rule. Specifically, we execute the test case against the seed bug program. If a test case fails, its execution trace often contains valuable information, such as the sequence of method calls and the statements involved when the bug occurred. 
We leverage an LLM to analyze the execution trace, evaluating whether the failure message and runtime behavior align with the expected characteristics of the bug as defined by the detection rule. If the alignment is weak or unclear, the test case undergoes iterative refinement for a predefined number of attempts. If the seed bug cannot be reliably triggered by a test case after multiple refinements, we discard the seed bug and generate a new one. This automated validation process ensures that only valid, reproducible bugs are retained, improving the accuracy and reliability of the dataset for static analysis evaluation. 

For the seed bug in Listing~\ref{code:seed}, a test generated by this agent is shown in Listing~\ref{code:test}. The test case named \textit{testShowBugWithPolygenelubricants}, which calls the buggy method \textit{showBug} with a valid input that triggers the bug. Notably, in our approach, all the test case names and inputs were generated by the LLM without any constraints in the prompt, as the name does not affect the test case's ability to trigger bugs. 

\begin{lstlisting}[style=mystyle, label={code:test}, caption={The test case generated by \textit{Code Validation Agent} to test the \texttt{showBug(String input)} method in Listing \ref{code:seed}. The absolute value of the hashCode becomes negative due to the bug in Figure \ref{fig:ruleExamp}, and hence, the assertion (\texttt{result >= 0)} fails.}]
public class HashCodeBugTest {
    @Test
    public void testShowBugWithPolygenelubricants() {
        HashCodeBug example = new HashCodeBug();
        String input = "polygenelubricants";
        int result = example.showBug(input);
        assertTrue("Expected non-negative hashcode, but got: " + result, result >= 0);
    }
}
\end{lstlisting}

\subsection{Mutation Generation Agent}
\label{sec:3.4}

\input{tab/mutant}

In this work, the seed bugs generated using \textit{Seed Generation Agent} are intentionally simple to ensure clarity and reproducibility. 
To evaluate static analyzers on more complex bug instances, we further leverage this agent to generate semantically equivalent mutants~\cite{tian2024large} with LLMs, diversifying the bug programs while preserving their core defect characteristics. 

We reuse a set of nine mutation operators that can be used to generate semantically equivalent methods from existing work~\cite{wei2022cocofuzzing,zhang2023statfier}. 
These mutation operations range from common refactoring, such as variable renaming, to more intrusive ones, such as adding unreachable branches.
Table~\ref{tab:opts} provides a summary of the ten mutation operators and how each operator can be utilized to modify a program. 
Notably, Statfier~\cite{zhang2023statfier} includes several transformation operators that we intentionally did not adopt, such as Extract Local Variable, Move Assignment, and other equivalent expression-level operators. The rationale is that these transformations, while semantically valid, typically do not result in observable differences in the compiled bytecode. As a result, they are less effective for evaluating static analysis tools that operate at the bytecode level, such as SpotBugs, Infer, Errorprone, and SonarQube. 

The following outlines the specific mutation generation process for each operator.

\noindent \textbf{(Op1) Dead store:} inserts an unused variable declaration of one primitive type (e.g., string, int, double, and long, etc.) to a randomly selected basic block in the program. Only one dead store is added in each transformation by this operator.

\noindent \textbf{(Op2) Obfuscating:} This transformation rewrites a numerical value or variable within a statement by introducing a random numerical addition and subtraction of the same value, preserving the original computation. For example, x = 1.0; may be mutated to x = 1.0 + 0.1 - 0.1; or x = 1.0 + 0 - 0;. If the program contains multiple numerical variables, one is randomly selected for transformation. This operator applies only to assignment, declaration, and return statements to maintain syntactic and semantic validity.

\noindent \textbf{(Op3) Duplication:} duplicates a randomly selected assignment statement and inserts it immediately after its current location. To avoid side effects, the applicable assignment statement is limited to those without using method invocation.

\noindent \textbf{(Op4 to Op7) Unreachable loops/branches:} inserts an unreachable loop or branch (including \textit{if} statement, \textit{for} statement, \textit{while} statement, and \textit{switch} statement) into a randomly selected basic block in the program. 
The condition of the inserted loop or branch is always false to make it unreachable. 

\noindent \textbf{(Op8) Renaming:} renames a local variable declared in a program. If there exist multiple variables, we randomly select one for the mutation. The new name of the variable will be in the form of [a-z].  

\noindent \textbf{(Op9) Equivalent do-while:} This mutation operator randomly replaces a \texttt{for}/\texttt{while} loop in the seed program with an equivalent \texttt{do-while} loop syntax. If no loop exists in the seed program, then no changes are made. 

Given a seed bug and a set of mutation operators, we first use an LLM to determine which operators are applicable to the given seed bug. Once suitable operators are identified, we instruct the LLM to generate $N$ distinct mutants for each applicable operator, ensuring diversity in the transformations. 
Similar to the seed bug generation, the mutants may contain syntax errors due to LLM limitations in enforcing language constraints. To address this, we apply a compilation check to validate the syntactic correctness of each generated mutant. If a mutant fails to compile, we use an iterative refinement process, feeding both the error message and the mutant code back into the LLM for correction. This process is repeated until a valid, compilable mutant is obtained or a predefined attempt limit is reached, at which point the mutant is discarded, and a new one is generated. 

Another key constraint for the generated mutants is that they must be semantically equivalent to their corresponding seed bugs. This means they should preserve the original program's behavior while introducing structural variations. 
To validate semantic equivalence, we execute the same set of test cases on both the seed bug and its mutants. 
If the test results, i.e., output, exceptions, and execution traces, are identical, the mutant is considered a valid semantically equivalent variant. 
If discrepancies arise, we feed the differing outputs from the seed bug and its mutant back into the LLM, prompting it to refine the mutant to resolve the inconsistency. This iterative refinement process continues for a predefined number of attempts (e.g., up to five iterations). If the mutant remains semantically non-equivalent after multiple refinements, it is discarded, ensuring that only valid, behavior-preserving mutants are retained for further evaluation. Such a mutant of the seed bug from Listing \ref{code:seed} is shown in Listing \ref{code:mutant}, in which the mutation operator Op5 was applied and an unreachable \textit{switch} statement was inserted into the seed bug.


\begin{lstlisting}[style=mystyle, label={code:mutant}, caption={\small{Mutant (using OP5) generated by the \textit{Mutation Generation Agent} for the seed in Listing \ref{code:seed}. It adds a switch statement which falls back to \texttt{default} case and hence does not change the behavior.}}]
class HashCodeBug {
    public int showBug(String input) {
        int hashcode = input.hashCode();
        int absHashcode = Math.abs(hashcode);
        int unreachableValue = Integer.MIN_VALUE;
        switch (unreachableValue) {
            case 1:
                System.out.println("Unreachable case.");
                break;
            default:
                break;
        }
        return absHashcode;
    }
}
\end{lstlisting}

\subsection{Analyzer Evaluation Agent}
\label{sec:3.5}
This agent takes the seed bug and its corresponding semantically equivalent mutants to perform metamorphic testing on a specific bug detection rule within a static analyzer.  
Specifically, it calls a static analyzer to detect bugs in a seed bug program and its mutants. If the analyzer detects the seed bug but fails to identify some or all of its mutants, it suggests that the rule is overly specific and lacks robustness. If the analyzer misses both the seed bug and its semantically equivalent mutants, it suggests that the bug detection rule is fundamentally inadequate. 
This approach helps identify potential inconsistencies in the rule's implementation. We focus on detecting two types of issues:

\begin{itemize}[leftmargin=*]
\item \textbf{Type1: Inconsistent Detection}. The static analyzer detects the seed bug, but not all of its semantically equivalent mutants. 
 
\item \textbf{Type2: False Negatives Across Variants}. The static analyzer fails to detect both the seed bug and at least one of its mutants. 
\end{itemize}

The seed bug listed in Listing \ref{code:seed} and its mutant listed in Listing~\ref{code:mutant} trigger a \textit{Type1} error of SpotBugs.  
Specifically, the seed program computes the absolute value of the \texttt{input.hashCode()} value. However, calculating the absolute of hashCode values is error-prone. Because when the hash value is equal to \texttt{Integer.MIN\_VALUE}, the \texttt{Math.abs} operation on it becomes \texttt{Integer.MAX\_VALUE+1}. This overflows the data type and becomes a negative value again. SpotBugs detected this issue in the seed bug in Listing \ref{code:seed}. However, its simple mutation in Listing \ref{code:mutant} evaded detection by SpotBugs. Although the \texttt{Math.abs} operation was performed on the hashcode variable, the presence of an unreachable switch-statement made SpotBugs fail to detect this issue. This is an example of our \textit{Type1} bug where the static analyzer's detection is inconsistent between the seed and its valid mutants.


%% file: fig/overview.tex
\begin{figure*}[!t]
    \centering
    \includegraphics[width=\textwidth]{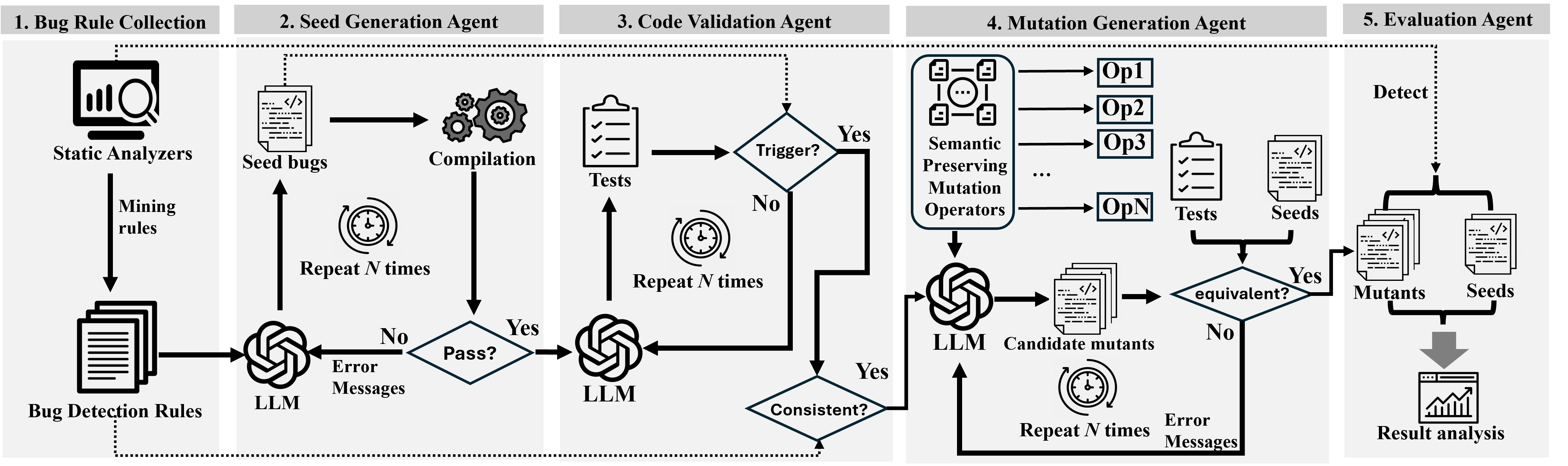}
    
    \caption{The workflow of \tool{}}
    \label{fig:overview}
\end{figure*}

%% file: fig/spotbugs_19.tex
\begin{figure}[!t]
    \centering
    \fbox{\includegraphics[width=0.44\textwidth]{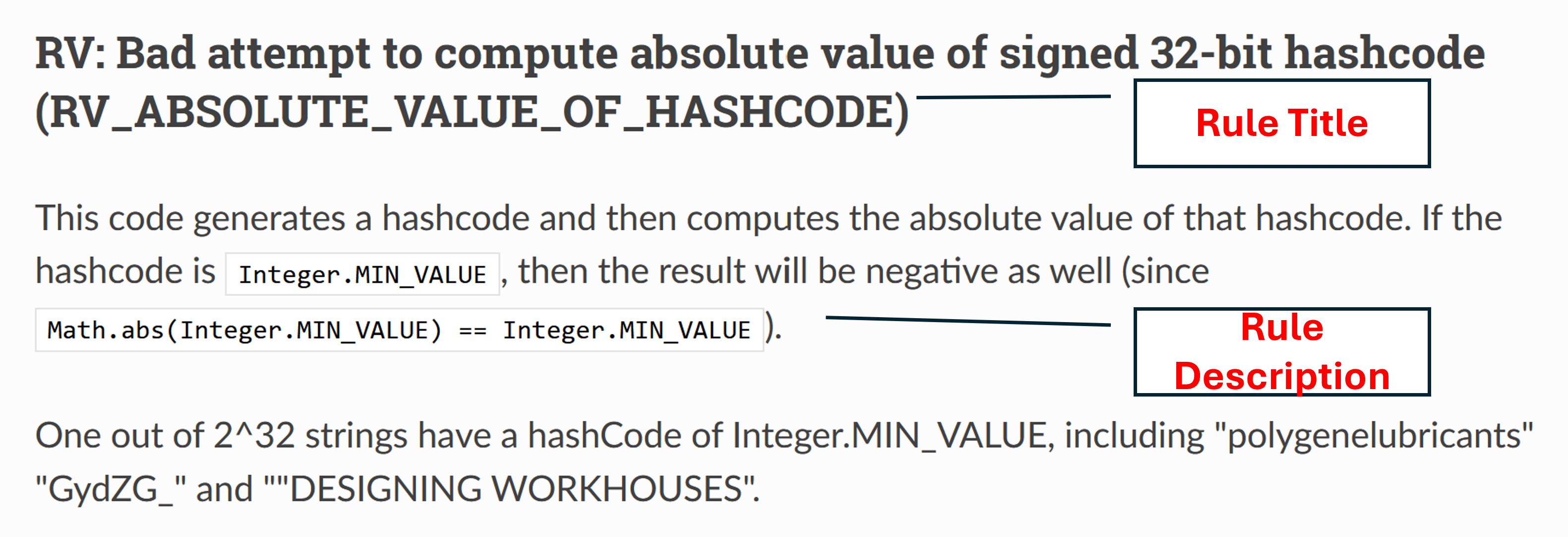}}
    \vspace{-0.1in}
    \caption{An example bug detection rule from Spotbugs}
    \label{fig:ruleExamp}
\end{figure}

%% file: tab/mutant.tex
\begin{table}[t!]
\centering
\caption{The semantic-preserving mutation operators from~\cite{wei2022cocofuzzing, zhang2023statfier}.}
\vspace{-0.1in}
\label{tab:opts}
\setlength{\tabcolsep}{3pt}
\scalebox{0.95}{
\begin{tabular}{lll}
\toprule 
\textbf{NO.} & \textbf{Operator name}  & \textbf{Description}  \\ \hline
Op1 & dead store            & Inserting unused variable declarations\\ \hline


Op2 & duplication          & Duplicating assignment statements\\ \hline
Op3 & unreachable \textit{if}   & Inserting unreachable \textit{if} statements \\ 
Op4 & unreachable \textit{if-else}  & Inserting unreachable \textit{if-else} statements \\ 
Op5 & unreachable \textit{switch}  & Inserting unreachable  \textit{switch} statements \\ 
Op6 & unreachable \textit{for}      & Inserting unreachable  \textit{for} statements \\ 
Op7 & unreachable \textit{while}   & Inserting unreachable  \textit{while} statements \\ \hline
Op8 & renaming & Renaming user-defined variables \\ \hline
Op9 & equivalent \textit{do-while} & Replace a for/while loop with \textit{do-while} \\\bottomrule \hline
\end{tabular}
}
\end{table}

%% file: sections/experiment.tex
\section{Experiment Design}
\label{sec:exp}

\subsection{\textbf{Subjective Static Analyzers Selection}}
In this study, we examined the correctness of static analysis rules in five static bug finders designed for Java: \textbf{SpotBugs}\footnote{\url{https://spotbugs.github.io/}}, \textbf{SonarQube}\footnote{\url{https://www.sonarsource.com/}}, \textbf{ErrorProne}\footnote{\url{https://errorprone.info/}}, \textbf{Infer}\footnote{\url{https://fbinfer.com/}}, and \textbf{PMD}\footnote{\url{https://pmd.github.io/}}. We chose these five tools based on their widespread use in research and industry adoption (3.7k--15.3k GitHub stars). All tools except PMD work on build artifacts, while PMD directly scans the source code.  
Note that some other popular tools, such as Checkstyle, were excluded because they primarily focus on style checking rather than detecting semantic or logical bugs.

\input{tab/rules}
\input{tab/results-new}

Table~\ref{tab:rules} shows the number of rules we extracted from each analyzer. The candidate rule counts considered for our study were 490 for SpotBugs, 704 for SonarQube, 594 for ErrorProne, 162 for Infer, and 291 for PMD. The candidate rules encompassed a wide range of Java source code analysis checks. We then refined the list to select rules that are directly related to critical aspects such as code correctness, security vulnerabilities, performance, and malware detection (details are in Section~\ref{sec:3.1}). Our study analyzed 199 rules for SpotBugs, 173 for SonarQube, 177 for ErrorProne, 26 for Infer, and 125 for PMD. Note that the number of selected rules for Infer is lower than for the other tools, primarily because many of Infer’s rules are either undocumented, too abstract, or tied to internal analysis constructs, making them difficult to extract and evaluate systematically.

\subsection{Experimental LLMs for Building {\tool}}

We selected five state-of-the-art LLMs to build our agentic framework, using the same LLM to power all four agents within each evaluation setting. Specifically, we select \emph{CodeLlama-34B-Instruct-hf}~\cite{roziere2023code}, \emph{Codestral-22B-v0.1}~\cite{jiang2023mistral}, \emph{Deepseek-Coder-33B-Instruct}~\cite{guo2024deepseek}, \emph{GPT-4o}~\cite{achiam2023gpt}, and \emph{Qwen2.5-Coder-32B}~\cite{hui2024qwen2} because these models are widely used in the community and have demonstrated high performance in various software engineering tasks. 

Note that when using LLMs to generate seed bugs, we rely on a basic zero-shot prompting approach. The primary reason for this choice is that not all static analysis rules come with example code, making it challenging to provide reference implementations. Additionally, manually crafting example code for each rule would be highly time-consuming and impractical, especially when dealing with a large number of rules across different analyzers. By using a zero-shot prompt, we ensure a scalable and automated approach to bug synthesis while still evaluating the LLM's ability to generalize and generate meaningful seed bugs based solely on the rule descriptions. The detailed prompts we used in each agent can be found in our replication package. 


\subsection{Experimental Setup}

We use all our models with different $temperature$ values based on the task. For generation tasks (seed, test, mutant), $temperature$ is set to 0.75, as it is a common choice for code generation tasks~\cite{gatto2024context, wu2024progress}. For validation tasks (seed and mutant validation), we use a lower value $temperature=0.1$ to ensure consistent and deterministic output. 
Other parameters were set to default values. Our experiments were carried out in a GPU cluster of 4 NVIDIA L40S along with Intel(R) Xeon(R) Gold 6442Y CPU. 

Note that {\tool} follows an interactive refinement process. In each iteration, we rerun the LLMs to generate compilable programs by leveraging error messages from the previous run. To optimize computational resources, we set the maximum number of refinement iterations to five across all interactive processes in our approach. For each seed bug, we generate three mutants per mutation operator. This number is chosen to balance diversity and efficiency, ensuring a sufficient variety of mutations while keeping the computational overhead manageable. 




%% file: tab/rules.tex
\begin{table}[]
\caption{The number of rules extracted from each analyzer.}
\vspace{-0.1in}
\label{tab:rules}
\begin{tabular}{l|c|c|c}
\hline
Analyzer  &Version&
\#Candidate Rules & \#Rules Selected \\ \hline
SpotBugs & 4.9.3 &  490  &  199    \\ \hline
SonarQube&  25.5 &  704   &   173 \\ \hline
ErrorProne & 2.40.0 & 594  &  177 \\ \hline 
Infer& 1.2.0 &  162  &   26 \\ \hline
PMD & 7.14.0 & 291 & 125 \\ \hline
\end{tabular}
\end{table}

%% file: tab/results-new.tex
\begin{table*}[t!]
\caption{Overall performance of different LLMs on different analyzers. `\#Comp Seed' denotes the number of compilable seeds among the generated ones. `\#Valid Seed' is the number of seed bugs that can be triggered by tests generated. `\#Comp Mutant' is the number of compilable mutants. `\#Valid Mutant' shows the number of semantics equivalent mutants. ``Buggy Rule Detection'' shows the number of rule implementation bugs for \#\textit{Type1} and \#\textit{Type2}, and the values denote LLM-detected bugs with manually verified true positives in parentheses.}

\setlength{\tabcolsep}{3pt}
\label{tab:performance-new}
\resizebox{\textwidth}{!}{
\begin{tabular}{ll|lll|lll|llll|}
\cline{3-12}
  & & \multicolumn{3}{c|}{Seeds and Tests Generation}& \multicolumn{3}{c|}{Mutant Generation}& \multicolumn{4}{c|}{Buggy Rule Detection}\\ \hline
\multicolumn{1}{|l|}{\textbf{Analyzer}}   & \textbf{Models} & \multicolumn{1}{l|}{\textbf{\#Comp Seeds}} & \multicolumn{1}{l|}{\textbf{\#Tests}} & \textbf{\#Valid Seeds} & \multicolumn{1}{l|}{\textbf{\#Mutants}} & \multicolumn{1}{l|}{\textbf{\#Comp Mutant}} & \textbf{\#Valid Mutant} & \multicolumn{1}{l|}{\textbf{\#\textit{Type1} }} & \multicolumn{1}{l|}{\textbf{FP \%}} & \multicolumn{1}{l|}{\textbf{\#\textit{Type2} }} & \textbf{FP \%} \\ \hline

\multicolumn{1}{|l|}{\multirow{5}{*}{\shortstack{SpotBugs \\ (199 rules)}}}   & CodeLlama   & \multicolumn{1}{l|}{81}& \multicolumn{1}{l|}{78} & 31 & \multicolumn{1}{l|}{837} & \multicolumn{1}{l|}{414} & 242 & \multicolumn{1}{l|}{2 ( 1 )} & \multicolumn{1}{l|}{50.00\%} & \multicolumn{1}{l|}{3 ( - )} & \multicolumn{1}{l|}{100\%} \\ \cline{2-12} 

\multicolumn{1}{|l|}{}& DeepSeek& \multicolumn{1}{l|}{195}& \multicolumn{1}{l|}{182} & \textbf{166} & \multicolumn{1}{l|}{\textbf{4482}} & \multicolumn{1}{l|}{3826} & 2671 & \multicolumn{1}{l|}{35 ( 5 )} & \multicolumn{1}{l|}{85.71\%}  & \multicolumn{1}{l|}{4 ( 1 )} & 75.00\% \\ \cline{2-12} 

\multicolumn{1}{|l|}{}& Codestral   & \multicolumn{1}{l|}{195}& \multicolumn{1}{l|}{\textbf{185}} & 160 & \multicolumn{1}{l|}{4320} & \multicolumn{1}{l|}{\textbf{4047}} & \textbf{3508} & \multicolumn{1}{l|}{23 (12)}& \multicolumn{1}{l|}{47.83\%} & \multicolumn{1}{l|}{6 ( 3 )} & \multicolumn{1}{l|}{50.00\%} \\ \cline{2-12}

\multicolumn{1}{|l|}{}& Qwen& \multicolumn{1}{l|}{193}& \multicolumn{1}{l|}{181} & 118 & \multicolumn{1}{l|}{3186} & \multicolumn{1}{l|}{2462} & 2251 & \multicolumn{1}{l|}{15 ( 10 )} & \multicolumn{1}{l|}{33.33\%}   & \multicolumn{1}{l|}{6 ( 2 )} & \multicolumn{1}{l|}{66.67\%} \\ \cline{2-12} 

\multicolumn{1}{|l|}{}& GPT-4o  & \multicolumn{1}{l|}{\textbf{197}}& \multicolumn{1}{l|}{183} & 117 & \multicolumn{1}{l|}{3159} & \multicolumn{1}{l|}{3134} & 2685 & \multicolumn{1}{l|}{7 ( 6 )} & \multicolumn{1}{l|}{\textbf{14.28\%}}   & \multicolumn{1}{l|}{5 ( 2 )} & \multicolumn{1}{l|}{60.00\%} \\ \hline \hline

\multicolumn{1}{|l|}{\multirow{5}{*}{\shortstack{SonarQube \\ (173 rules)}}} & CodeLlama & \multicolumn{1}{l|}{65} & \multicolumn{1}{l|}{88} & 48 & \multicolumn{1}{l|}{891} & \multicolumn{1}{l|}{555} & 318 & \multicolumn{1}{l|}{1 (-)} & \multicolumn{1}{l|}{100.00\%} & \multicolumn{1}{l|}{0 (-)} & \multicolumn{1}{l|}{0.00\%} \\ \cline{2-12}

\multicolumn{1}{|l|}{} & DeepSeek & \multicolumn{1}{l|}{165} & \multicolumn{1}{l|}{144} & \textbf{131} & \multicolumn{1}{l|}{4023} & \multicolumn{1}{l|}{3372} & 1973 & \multicolumn{1}{l|}{25 (3)} & \multicolumn{1}{l|}{88\%} & \multicolumn{1}{l|}{10(1)} & \multicolumn{1}{l|}{90.00\%} \\ \cline{2-12}

\multicolumn{1}{|l|}{} & Codestral & \multicolumn{1}{l|}{\textbf{167}} & \multicolumn{1}{l|}{149} & 120 & \multicolumn{1}{l|}{\textbf{4509}} & \multicolumn{1}{l|}{4120} & \textbf{2236} & \multicolumn{1}{l|}{19 (5)} & \multicolumn{1}{l|}{\textbf{73.68\%}} & \multicolumn{1}{l|}{21 (2)} & \multicolumn{1}{l|}{90.48\%} \\ \cline{2-12}

\multicolumn{1}{|l|}{} & Qwen & \multicolumn{1}{l|}{164} & \multicolumn{1}{l|}{\textbf{151}} & 85 & \multicolumn{1}{l|}{4428} & \multicolumn{1}{l|}{3476} & 544 & \multicolumn{1}{l|}{4 (1)} & \multicolumn{1}{l|}{75.00\%} & \multicolumn{1}{l|}{6 (2)} & \multicolumn{1}{l|}{66.67\%} \\ \cline{2-12}

\multicolumn{1}{|l|}{} & GPT-4o & \multicolumn{1}{l|}{165} & \multicolumn{1}{l|}{150} & 84 & \multicolumn{1}{l|}{4452} & \multicolumn{1}{l|}{\textbf{4295}} & 2109 & \multicolumn{1}{l|}{4 (1)} & \multicolumn{1}{l|}{75.00\%} & \multicolumn{1}{l|}{20 (3)} & \multicolumn{1}{l|}{85.00\%} \\ \hline \hline

\multicolumn{1}{|l|}{\multirow{5}{*}{\shortstack{ErrorProne\\ (177 rules)}}} & CodeLlama   & \multicolumn{1}{l|}{67}& \multicolumn{1}{l|}{62} & 28 & \multicolumn{1}{l|}{756} & \multicolumn{1}{l|}{379} & 235 & \multicolumn{1}{l|}{1 (-)}  & \multicolumn{1}{l|}{100.00\%}  & \multicolumn{1}{l|}{6 (-)}  & \multicolumn{1}{l|}{100.00\%} \\ \cline{2-12} 

\multicolumn{1}{|l|}{}& DeepSeek & \multicolumn{1}{l|}{164}& \multicolumn{1}{l|}{157} & \textbf{131} & \multicolumn{1}{l|}{\textbf{3537}} & \multicolumn{1}{l|}{\textbf{3003}} & 2172 & \multicolumn{1}{l|}{13 (1)}   & \multicolumn{1}{l|}{92.31\%} & \multicolumn{1}{l|}{9 (-)} & \multicolumn{1}{l|}{100.00\%} \\ \cline{2-12} 

\multicolumn{1}{|l|}{}& Codestral & \multicolumn{1}{l|}{\textbf{167}}& \multicolumn{1}{l|}{154} & 119 & \multicolumn{1}{l|}{3213} & \multicolumn{1}{l|}{2950} & \textbf{2346} & \multicolumn{1}{l|}{4 (2)}   & \multicolumn{1}{l|}{50.00\%} & \multicolumn{1}{l|}{1 (1)} & \multicolumn{1}{l|}{0.00\%} \\ \cline{2-12}

\multicolumn{1}{|l|}{}& Qwen & \multicolumn{1}{l|}{160}& \multicolumn{1}{l|}{144} & 78 & \multicolumn{1}{l|}{2106} & \multicolumn{1}{l|}{1413}& 1267 & \multicolumn{1}{l|}{3 (2)}   & \multicolumn{1}{l|}{\textbf{33.33\%}} & \multicolumn{1}{l|}{1 (1)} & \multicolumn{1}{l|}{0.00\%} \\ \cline{2-12}

\multicolumn{1}{|l|}{}& GPT-4o  & \multicolumn{1}{l|}{171}& \multicolumn{1}{l|}{\textbf{160}} & 84 & \multicolumn{1}{l|}{2268} & \multicolumn{1}{l|}{2138} & 1891 & \multicolumn{1}{l|}{0 (-)} & \multicolumn{1}{l|}{0.00\%} & \multicolumn{1}{l|}{4 (2)} & \multicolumn{1}{l|}{50.00\%} \\ \hline \hline

\multicolumn{1}{|l|}{\multirow{5}{*}{\shortstack{Infer\\ (26 rules)}}} & CodeLlama   & \multicolumn{1}{l|}{25}& \multicolumn{1}{l|}{25} & 10 & \multicolumn{1}{l|}{627} & \multicolumn{1}{l|}{349} & 158 & \multicolumn{1}{l|}{0 (-)}  & \multicolumn{1}{l|}{0.00\%}  & \multicolumn{1}{l|}{0 (-)} & \multicolumn{1}{l|}{0.00\%} \\ \cline{2-12}

\multicolumn{1}{|l|}{}& DeepSeek & \multicolumn{1}{l|}{\textbf{26}}& \multicolumn{1}{l|}{\textbf{26}} & \textbf{24} & \multicolumn{1}{l|}{\textbf{702}} & \multicolumn{1}{l|}{629} & 330 & \multicolumn{1}{l|}{4 (-)}  & \multicolumn{1}{l|}{100.00\%}  & \multicolumn{1}{l|}{2 (2)} & \multicolumn{1}{l|}{0.00\%} \\ \cline{2-12}

\multicolumn{1}{|l|}{}& Codestral & \multicolumn{1}{l|}{\textbf{26}}& \multicolumn{1}{l|}{24} & 17 & \multicolumn{1}{l|}{\textbf{702}} & \multicolumn{1}{l|}{670} & \textbf{386} & \multicolumn{1}{l|}{1 (-)}  & \multicolumn{1}{l|}{100.00\%}  & \multicolumn{1}{l|}{4 (-)} & \multicolumn{1}{l|}{100.00\%} \\ \cline{2-12}

\multicolumn{1}{|l|}{}& Qwen & \multicolumn{1}{l|}{\textbf{26}}& \multicolumn{1}{l|}{26} & 9 & \multicolumn{1}{l|}{\textbf{702}} & \multicolumn{1}{l|}{642} & 231 & \multicolumn{1}{l|}{0 (-)}  & \multicolumn{1}{l|}{0.00\%}  & \multicolumn{1}{l|}{2 (1)} & \multicolumn{1}{l|}{50.00\%} \\ \cline{2-12}

\multicolumn{1}{|l|}{}& GPT-4o  & \multicolumn{1}{l|}{\textbf{26}}& \multicolumn{1}{l|}{24} & \textbf{7} & \multicolumn{1}{l|}{\textbf{702}} & \multicolumn{1}{l|}{\textbf{689}} & 213 & \multicolumn{1}{l|}{0 (-)} & \multicolumn{1}{l|}{0.00\%} & \multicolumn{1}{l|}{4 (1)} & \multicolumn{1}{l|}{75\%} \\ \hline \hline

\multicolumn{1}{|l|}{\multirow{5}{*}{\shortstack{PMD\\ (125 rules)}}} & CodeLlama   & \multicolumn{1}{l|}{48}& \multicolumn{1}{l|}{38} & 35 & \multicolumn{1}{l|}{945} & \multicolumn{1}{l|}{543} & 268 & \multicolumn{1}{l|}{0 (-)}  & \multicolumn{1}{l|}{0.00\%}  & \multicolumn{1}{l|}{5 (1)} & \multicolumn{1}{l|}{80.00\%} \\ \cline{2-12} 

\multicolumn{1}{|l|}{}& DeepSeek & \multicolumn{1}{l|}{120}& \multicolumn{1}{l|}{116} & \textbf{98} & \multicolumn{1}{l|}{\textbf{2646}} & \multicolumn{1}{l|}{\textbf{2244}} & 1103 & \multicolumn{1}{l|}{16 (4)}  & \multicolumn{1}{l|}{75.00\%}  & \multicolumn{1}{l|}{14 (-)} & \multicolumn{1}{l|}{100.00\%}\\ \cline{2-12} 

\multicolumn{1}{|l|}{}& Codestral   &  \multicolumn{1}{l|}{110}& \multicolumn{1}{l|}{109} & 69 & \multicolumn{1}{l|}{1863} & \multicolumn{1}{l|}{1652} & \textbf{1318} & \multicolumn{1}{l|}{11 (6)}  & \multicolumn{1}{l|}{45.45\%}  & \multicolumn{1}{l|}{3 (-)} & \multicolumn{1}{l|}{100.00\%} \\ \cline{2-12} 

\multicolumn{1}{|l|}{} & Qwen & \multicolumn{1}{l|}{\textbf{125}}& \multicolumn{1}{l|}{\textbf{122}} & 50 & \multicolumn{1}{l|}{1350} & \multicolumn{1}{l|}{1085}& 833 & \multicolumn{1}{l|}{1 (1)} & \multicolumn{1}{l|}{\textbf{0.00\%}} & \multicolumn{1}{l|}{2 (-)} & \multicolumn{1}{l|}{100.00\%} \\ \cline{2-12} 

\multicolumn{1}{|l|}{}& GPT-4o  & \multicolumn{1}{l|}{124}& \multicolumn{1}{l|}{117} & 52 & \multicolumn{1}{l|}{1404} & \multicolumn{1}{l|}{1383} & 1088 & \multicolumn{1}{l|}{1 (-)} & \multicolumn{1}{l|}{100.00\%} & \multicolumn{1}{l|}{2 (-)} & \multicolumn{1}{l|}{100.00\%} \\ \hline
\end{tabular}}
\end{table*}

%% file: sections/results.tex
\section{Result Analysis}
\label{sec:results}


This section presents our experimental results and evaluates the performance of {\tool}. 


\subsection{Performance of \textit{Seed Generation Agent}}

Our pipeline begins by using \textit{Seed Generation Agent} to generate seed programs for each rule in an analyzer and replicating the bug described in the rule. 
For the seed bug generation task, most of the models perform convincingly well, except for CodeLlama. These models have compile success rates from 95.22\%--98.29\% across all five tools, where CodeLlama's compilation rate is only 40.92\%. Detailed breakdown of each model and each tool is given in \textbf{\#Comp Seeds} column in Table~\ref{tab:performance-new}. However, despite the high compile rate, the models fail to produce compilable code on certain occasions. One of the common problems is when the bug rule involves mixing different types of variables. When two incompatible types are compared with the pointer equal operator (\texttt{==}) or \texttt{instanceof} operator, certain problems may arise. For example, BC: BC\_IMPOSSIBLE\\\_INSTANCEOF\footnote{\url{https://spotbugs.readthedocs.io/en/stable/bugDescriptions.html\#bc-instanceof-will-always-return-false-bc-impossible-instanceof}} rule from SpotBugs checks if the \texttt{instanceof} operator is used incorrectly and always evaluates to false. In Listing \ref{code:not_compile_example}, we can see a seed generated for this rule by the Qwen2.5-Coder-32B model. In the program, a \texttt{String} variable was compared to an \texttt{Integer} with an \texttt{instanceof} operation. However, \texttt{String} and \texttt{Integer} are incompatible types, and line 4 leads to a compilation error. Some other common causes of compilation failures are incorrect method signatures while overriding, incorrect or non-existent annotation imports/usage, etc.

\begin{lstlisting}[style=mystyle, label={code:not_compile_example}, caption={A seed program generated by Qwen2.5 fails to compile due to an incompatible type error at line 4.}]
class ImpossibleInstanceof {
    public boolean showBug() {
        String str = "Hello, World!";
        boolean result = str instanceof Integer;
        System.out.println(result);
        return result;
    }
}
\end{lstlisting}

\mybox{The \textit{Seed Generation Agent}, when powered by all LLMs except CodeLlama, demonstrates strong effectiveness in producing compilable seed bugs, achieving compilation rates between 95.22\% and 98.29\%. Common causes of compilation failures include incorrect method overriding, incompatible type comparisons, and the use of invalid or non-existent annotations.}

\subsection{Performance of \textit{Code Validation Agent}}

The \textit{Code Validation Agent} focuses on generating test cases to confirm that the seed programs accurately contain the desired bugs, as specified by the rules. As in the seed generation phase, we first discard any tests that fail to compile; the count of compilable tests appears in the ``\#Test'' column in Table~\ref{tab:performance-new}. 
We then execute the remaining tests and employ LLMs to determine whether each test triggers the same bug described by its corresponding rule. The number of seeds confirmed as valid by the LLMs is reported in the ``\#Valid Seeds'' column in Table~\ref{tab:performance-new}.

\begin{figure}
\begin{tikzpicture}
    \begin{axis}[
     height=3.5cm,
    width=9cm,
        ybar,
        symbolic x coords={CodeLlama, Codestral, DeepSeek, GPT-4o, Qwen},
        xtick=data,
        ymin=0, ymax=1,
        ylabel={Accuracy},
          nodes near coords,
        bar width=15pt
    ]
        \addplot coordinates {(CodeLlama,0.3966) (Codestral,0.6131) (DeepSeek,0.5742) (GPT-4o,0.7912) (Qwen,0.7978)};
    \end{axis}
\end{tikzpicture}
    \caption{Accuracy of the ``Valid Seeds'' generated by each LLM}
    \label{fig:validSeed}
\end{figure}
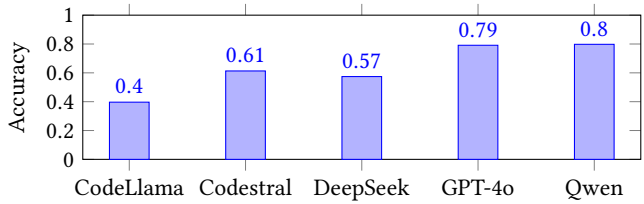

To evaluate seed validation accuracy, we manually examined the validated seed bugs from each of the five static analyzers for each LLM. Initially, we analyzed the seed bug programs along with the rule descriptions to determine whether these are truly buggy. This process established the ground truth for the analyzed seeds. We then evaluated each LLM’s accuracy by contrasting its judgments with this manually derived ground truth. We show the accuracy of valid seed bug generation for each LLM in Figure \ref{fig:validSeed}. 
It reveals that although DeepSeek labels many seeds as valid (``\#Valid Seeds'' in Table~\ref{tab:performance-new}), many of them fail to manifest the intended violation in practice. In contrast, GPT‑4o and Qwen, despite producing fewer valid seeds overall, achieve substantially higher validation precision. These findings indicate that test generation and validation by LLMs can suffer from significant false positives depending on the model. Exploring advanced prompting or Chain‑of‑Thought (CoT) strategies may improve performance for models such as CodeLlama and DeepSeek. 

A key challenge in the validation process arises from the inherent difficulty of generating tests for certain rules. For example, Listing~\ref{code:dead_thread} illustrates a seed program targeting the DeadThread\footnote{\url{https://errorprone.info/bugpattern/DeadThread}} rule in ErrorProne, which warns when a thread is instantiated but never started. Because Java does not throw an exception for an unused thread, the test in Listing~\ref{code:dead_thread_test} passes without error, and thus fails to detect the dead-thread bug present at line 3. 

\begin{lstlisting}[style=mystyle, label={code:dead_thread}, caption={A seed bug generated by DeepSeek-Coder-33B that demonstrates the creation of a dead thread.}]
class DeadThreadBug {
    public void showBug() {
        Thread t = new Thread(new Runnable() {
            public void run() {
                System.out.println("Thread running");}});
    }
}
\end{lstlisting}
\vspace{-\baselineskip}
\begin{lstlisting}[style=mystyle, label={code:dead_thread_test}, caption={The test case generated for the seed in Listing~\ref{code:dead_thread}.}]
public class DeadThreadBugTest {
    @Test
    public void testShowBug() {
        DeadThreadBug bug = new DeadThreadBug();
        bug.showBug();
    }
}
\end{lstlisting}

During our manual analysis, we also observed that LLMs sometimes misclassify genuinely buggy seeds as invalid. One common scenario occurs when the test wraps the bug-triggering code in a \texttt{try-catch} block or uses JUnit’s \texttt{expected} attribute in the \texttt{@Test} annotation, thereby suppressing the observable failure. Listing~\ref{code:null_expected_test} exemplifies this: because the test expects a \texttt{NullPointerException} at line 6, it passes when the exception is thrown, yet the LLM fails to recognize this complex situation that the seed is indeed defective.

\begin{lstlisting}[style=mystyle, label={code:null_expected_test}, caption={{A test case that passes a seed bug by expecting a NullPointerException can lead the LLM to incorrectly conclude that the seed contains no bug, as the test passes without triggering the expected issue.}}]
public class AsyncCallableTest {
    @Test(expected = NullPointerException.class)
    public void testShowBugTriggersNullPointerException() throws ExecutionException, InterruptedException {
        AsyncCallable example = new AsyncCallable();
        Future<Void> result = example.showBug();
        result.get();
    }}
\end{lstlisting}



\mybox{The \textit{Code Validation Agent}, when powered by GPT‑4o and Qwen, achieved the highest validation accuracy (~79\%). However, some seed programs were incorrectly validated due to flawed test generation or reasoning errors by the LLMs.}

\subsection{Performance of \textit{Mutation Generation Agent}}
\label{sec:5.3}

In this stage of the pipeline, the \textit{Mutation Generation Agent} generates mutant programs for each validated seed by applying a set of mutation operators. For every operator, we prompt an LLM to produce three semantically equivalent variations per seed program. 
The mutant programs went through the same steps as the seed programs. They were first compiled, and following successful compilation, they went through the same validation process (see Section~\ref{sec:3.3}). 
Similar to the seed bug generation, only compilable mutants are retained for validation to ensure they are semantically equivalent to the corresponding seed bugs. 

\begin{figure}
\begin{tikzpicture}
    \begin{axis}[
     height=3.5cm,
    width=9cm,
        ybar,
        symbolic x coords={Op1, Op2, Op3, Op4, Op5, Op6, Op7, Op8, Op9},
        xtick=data,
        ymin=0, ymax=1,
        ylabel={Percentage},
          nodes near coords,
        bar width=15pt
    ]
        \addplot coordinates {(Op1,.74) (Op2,.66) (Op3,.71) (Op4,.66) (Op5,.62) (Op6,.67) (Op7,.67) (Op8,.53) (Op9,.72)};
    \end{axis}
\end{tikzpicture}
    \caption{The percentage of semantically valid mutants generated per mutation operator}
    \label{fig:operator-valid}
\end{figure}
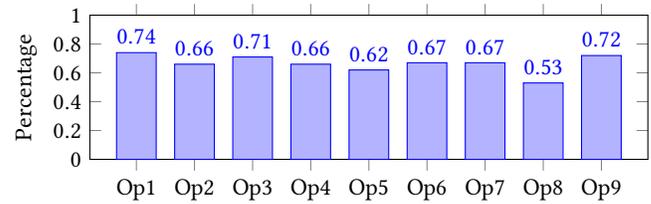

Figure~\ref{fig:operator-valid} shows the percentage of valid mutants for each mutation operator. We found that Op1 had the best performance, with 74\% of the mutants with Op1 being valid. This can be explained by Op1 being less intrusive compared to other mutant operators. 
In general, 68.86\% of the generated mutants were valid, providing ample data for metamorphic testing. 


We also manually verified that for the mutants that failed to compile, the issues stemmed from incorrect code structure, scope violations, 
name conflicts, or the introduction of code that disrupted the logical flow of the program. 
One of the common cases is the incorrect construction of unreachable branches. The LLMs tried to create unreachable loop mutants by introducing statements such as \texttt{while (false)} or \texttt{for (;false;)}. However, these loops will cause a compilation error in Java. Another case is where a method misses a return statement due to unreachable branching. The code in the \texttt{if (false)}-branch in Listing \ref{code:no_return_example} is unreachable, and the LLM did not put any code inside this branch. But since the method has a return type, something must be returned from all execution paths, including this if-branch. Therefore, the lack of a return statement results in a compilation error.
\begin{lstlisting}[style=mystyle, label={code:no_return_example}, caption={No return statement in if-branch leads to compile failure for this mutant. (Code is trimmed for limited space)}]
if (false) {
    // Unreachable code
} else {
    return output;
}
\end{lstlisting}
Some of the mutants that successfully compiled were also filtered out during the validation process. Some of the reasons for failing the validations are similar to the reasons for seed validation. However, DeepSeek and CodeLlama showed different behaviors from other models. When generating the mutants, they would occasionally fix the bug in the seed. Therefore, the mutant would no longer trigger the test case and would be ruled out of the validation. 

\mybox{Overall, the \textit{Mutation Generation Agent} is capable of producing valid mutants for the seed bugs, with 68.86\% of the generated mutants successfully passing validation.}



\subsection{Performance of \textit{Analyzer Evaluation Agent}}

\input{tab/uniqueBugs}





The \textit{Analyzer Evaluation Agent} conducts metamorphic testing by comparing the static analyzer’s behavior on seed programs and their corresponding mutants.
This section presents our key findings on incorrect or inconsistent rule implementations identified in the evaluated static analysis tools by this agent. 
The column ``Buggy Rule Detection'' in Table \ref{tab:performance-new} shows the number of \textit{Type1} and \textit{Type2}  (definitions are in Section \ref{sec:3.5}) rule violations detected across all five tools and all five models. 
We further show the unique bugs detected in each static analyzer in Table~\ref{tab:uniqueBugs}. We defined \emph{unique bug} as (i) bugs in different rules, (ii) bugs due to different root causes in the same rule.
For our analysis, two of the authors manually validated each bug instance detected by our pipeline and identified the true positive bugs. 

\subsubsection*{\textbf{Bug Detection Performance:}}
As shown in Table \ref{tab:performance-new} and Table~\ref{tab:uniqueBugs}, both Type 1 and Type 2 bugs were detected across all five analyzed tools. SpotBugs exhibited a higher number of incorrect rule implementations for both bug types, with 22 unique \textit{Type1} bugs and 6 unique \textit{Type2} bugs. Among the models evaluated, Codestral-22B detected the highest number of incorrect rule implementations, uncovering 15, 7, 3, and 6 bugs in SpotBugs, SonarQube, ErrorProne, and PMD, respectively. 
In addition to Codestral-22B, Qwen2.5 and GPT-4o also demonstrated strong performance, successfully identifying 20 and 15 bugs while maintaining low false-positive rates.


\subsubsection*{\textbf{Comparison to Baseline:}}
In this section, we compare \textsc{StaAgent}'s performance against the SOTA approach for detecting bugs in static analyzers, i.e., Statfier~\cite{zhang2023statfier}. 
For a fair comparison, we tested the latest versions of the analyzers using Statfier’s publicly available replication package\footnote{\url{https://github.com/cshezhang/Statfier}}. Except for five PMD rules that were deprecated in its most recent version, we retained the same set of seeds for all rules across all tools. Since Statfier does not support ErrorProne and therefore lacks corresponding seeds, our evaluation excludes ErrorProne from this comparison. 

Table~\ref{tab:uniqueBugs} presents the number of unique rule violations detected by both Statfier and \textsc{StaAgent}. \textsc{StaAgent} significantly outperforms Statfier on the latest versions of SpotBugs and SonarQube, detecting a greater number of bugs. Of the 12 bugs identified by Statfier for SpotBugs, 8 were also detected by \textsc{StaAgent}. Notably, Statfier did not find any new bugs specifically for SonarQube. We hypothesize that the SonarQube bugs previously reported in the original Statfier paper have been addressed in the latest release, and their seeds and mutations are not effective in exposing issues in the updated version. For PMD, Statfier detected 3 more bugs than \textsc{StaAgent}; however, \textsc{StaAgent} discovered 5 unique bugs that Statfier missed, demonstrating the complementary strengths of the two approaches.


\subsubsection*{\textbf{False-Positive Rates:}}
Despite the successful detection of many buggy rules, some false positives were also generated, especially for \textit{Type2}. The DeepSeek model notably exhibited a high false-positive rate, as detailed in Table~\ref{tab:performance-new}. This issue primarily stemmed from two factors. First, DeepSeek sometimes inadvertently corrected bugs during mutant generation, resulting in mutants not flagged by the analyzers and thus triggering false \textit{Type1} alarms. Second, during test generation, DeepSeek sometimes rewrote the test code instead of consistently invoking the predefined \texttt{showBug()} method as instructed in the prompt. This unintended modification could inadvertently introduce a bug into the test, causing it to fail even if the original seed file was bug-free. Consequently, although the seed file remained correct and undetected by the analyzers, the test failed for the buggy mutant, producing \textit{Type2} false positives. These two factors significantly inflated DeepSeek's reported alerts, with most being false positives. Additionally, CodeLlama proved to be ineffective for this task due to its low generation rate of valid seeds.

Beyond model-specific issues, some common scenarios contributed to false positives across all models. A key challenge was the incorrect implementation of intended bugs in seed programs. For instance, the SpotBugs rule COMPARING\_INCOMPATIBLE\_TYPE\\\_QUALIFIERS\footnote{https://spotbugs.readthedocs.io/en/stable/bugDescriptions.html} requires particular annotations to manifest. However, the LLMs often generated seed programs without correctly using or importing the necessary annotations. 
Another contributing factor was the use of \texttt{try-catch} blocks around buggy statements. For certain exceptions (e.g., NullPointerException), if the exception was already handled and the \texttt{catch} block did not produce new exceptions, the tools would not report a detection. 


\mybox{{\tool} identified 43 \textit{Type1} and 21 \textit{Type2} faulty rule implementation bugs in the five tools. Codestral, Qwen2.5, and GPT-4o showed strong performance, while DeepSeek and CodeLlama were less effective and proved unsuitable for the task.}

\section{Discussion}
\label{sec:dis}
\subsection{Distribution of Detected Bugs by Mutation Operator}
\begin{figure}[t]
\begin{tikzpicture}
    \begin{axis}[
     height=4cm,
    width=9cm,
        ybar,
        symbolic x coords={Op1, Op2, Op3, Op4, Op5, Op6, Op7, Op8, Op9},
        xtick=data,
        ymin=0, ymax=22,
        ylabel={The number of bugs},
        bar width=15pt
    ]
        \addplot coordinates {(Op1,5) (Op2,6) (Op3,17) (Op4,16) (Op5,13) (Op6,17)(Op7,20) (Op8,6) (Op9,7)};
    \end{axis}
\end{tikzpicture}
    \caption{The number of \textit{Type1} bugs across the analyzers detected by applying each mutation operator.}
    \label{fig:operator-contribution}
\end{figure}
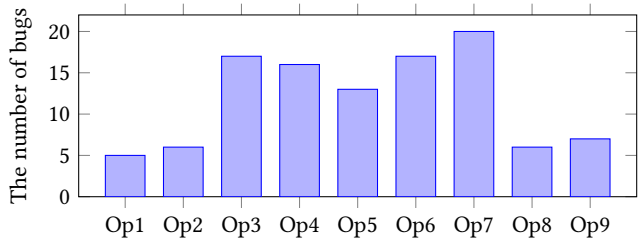

We further examined the effectiveness of various mutation operators in revealing \textit{Type1} bugs within our testing framework. Figure~\ref{fig:operator-contribution} illustrates the number of defects identified by the mutants generated by each operator.

As illustrated in Figure~\ref{fig:operator-contribution}, overall, all the mutation operators are useful in helping {\tool} detect Type 1 bugs in static analyzers. The number of bugs detected ranges from five to 20. 
Mutations that significantly alter the control flow tend to expose more hidden bugs. Specifically, our mutation operators Op3 through Op7 introduce complex branching constructs such as \texttt{if-else}, \texttt{switch-case}, and \texttt{for}/\texttt{while} loops into the seed programs, thereby increasing control-flow complexity. This added complexity poses greater challenges to static analyzers and is more effective in uncovering flaws in rule implementations. In contrast, mutations such as dead store insertion (Op1), duplicate statements (Op2), or variable renaming (Op8) minimally impact the program's control flow or structural semantics, limiting their ability to reveal inconsistencies.


\subsection{\textsc{StaAgent} on Seeds from Statfier}

By default, \textsc{StaAgent} relies on rule descriptions to validate rule implementations. In this section, we present additional experiments to evaluate the effectiveness of \textsc{StaAgent} in scenarios where seed programs are already available. For this purpose, we used the seed programs provided by Statfier~\cite{zhang2023statfier} as input. Assuming these seeds to be valid, we excluded the Seed Generation and Code Validation agents from the \textsc{StaAgent} pipeline, while keeping the remaining stages unchanged. We conducted this experiment on SpotBugs and Infer. ErrorProne was excluded as Statfier did not include it and thus provided no seeds. For PMD, many Statfier seeds were uncompilable due to undeclared identifiers, which is acceptable for PMD’s source-level analysis but incompatible with our pipeline, which requires executable tests. SonarQube was also excluded due to differences in project structure, rule tagging, and server configuration between our setup and Statfier’s. SpotBugs and Infer were selected as they support file-level scanning and allow a fair comparison. We conducted these experiments using the Qwen2.5-Coder-32B, as it offers a favorable trade-off between substantial bug discovery and minimal false-positive rates. Note that this experiment can only detect Type1 bugs, as the seeds are derived from the analyzers’ own test suites and are, by default, detectable by the corresponding tools. 

In total, 339 Infer seeds and 868 SpotBugs seeds were used for mutation generation. For Infer, only 56 seeds successfully triggered rule violations; this may be due to Infer's reliance on specific conditions such as annotations, special libraries, or flags. In contrast, 491 SpotBugs seeds triggered rule violations; the remaining seeds require specific execution contexts that the seed programs did not satisfy.  Table~\ref{tab:statfier_seeds} shows the performance of {\tool} in identifying \emph{Type1} bugs compared to Statfier using the \emph{seeds from Statfier}. {\tool} identified two bugs in Infer and ten bugs in SpotBugs. We manually verified that these bugs are distinct from those reported by Statfier, which could be attributed to the different mutation generation strategies employed by {\tool} and Statfier. This highlights the effectiveness of {\tool} in uncovering additional bugs through generalized seed programs. Overall, {\tool} and Statfier are complementary in their ability to detect diverse sets of bugs in static analyzers.

\input{tab/statfier_seeds}

%% file: tab/uniqueBugs.tex
\begin{table}[t!]
\centering
\caption{The number of \textbf{unique bugs} detected on each tool, combining all LLMs. \textbf{Common} column shows the number of unique bugs detected by both approaches.}
\vspace{-0.1in}
\label{tab:uniqueBugs}
\setlength{\tabcolsep}{3pt}
\begin{tabular}{l|rccc|c|}
\cline{2-6}  & \multicolumn{3}{c||}{\textbf{\textsc{StaAgent}}} & \multirow{2}{*}{\textbf{Statfier}} & \multirow{2}{*}{\textbf{Common}} \\ 
\cline{2-4} 
       & \multicolumn{1}{l|}{\#\textit{Type1}} & \multicolumn{1}{c|}{\#\textit{Type2}} & \multicolumn{1}{c||}{Total} & \multicolumn{1}{c|}{} & \\ \hline
\multicolumn{1}{|l|}{SpotBugs} & \multicolumn{1}{c|}{22}  & \multicolumn{1}{c|}{6} & \multicolumn{1}{c||}{\textbf{28}} & \multicolumn{1}{c|}{12} & \multicolumn{1}{c|}{8} \\ \hline
\multicolumn{1}{|l|}{SonarQube} & \multicolumn{1}{c|}{10} &  \multicolumn{1}{c|}{8} & \multicolumn{1}{c||}{\textbf{18}} & \multicolumn{1}{c|}{0} & \multicolumn{1}{c|}{0} \\ \hline
\multicolumn{1}{|l|}{ErrorProne} & \multicolumn{1}{c|}{4} & \multicolumn{1}{c|}{2} & \multicolumn{1}{c||}{\textbf{6}} & \multicolumn{1}{c|}{-} & \multicolumn{1}{c|}{-} \\ \hline
\multicolumn{1}{|l|}{Infer} & \multicolumn{1}{c|}{0} & \multicolumn{1}{c|}{4} & \multicolumn{1}{c||}{\textbf{4}} & \multicolumn{1}{c|}{2} & \multicolumn{1}{c|}{0} \\ \hline
\multicolumn{1}{|l|}{PMD} & \multicolumn{1}{c|}{7} & \multicolumn{1}{c|}{1} & \multicolumn{1}{c||}{8} & \multicolumn{1}{c|}{\textbf{11}} & \multicolumn{1}{c|}{3} \\ \hline \hline
\multicolumn{1}{|l|}{\textbf{Overall}}  & \multicolumn{1}{c|}{43} & \multicolumn{1}{c|}{21} & \multicolumn{1}{c||}{64} & \multicolumn{1}{c|}{25} & \multicolumn{1}{c|}{11} \\ \hline
\end{tabular}
\end{table}

%% file: tab/statfier_seeds.tex
\begin{table}[t!]
\centering
\caption{The number of Type1 bugs detected with Statfier's seeds by Statfier and StaAgent.}
\label{tab:statfier_seeds}
\setlength{\tabcolsep}{3pt}
\resizebox{0.5\textwidth}{!}{
\begin{tabular}{lccc|cc|c|}
\cline{3-7} & & \multicolumn{2}{|c|}{StaAgent} & \multicolumn{2}{c|}{Statfier} & Common \\
\cline{2-6}
& \multicolumn{1}{|l|}{\#Seeds} & \multicolumn{1}{|l|}{\#Mutants} & \multicolumn{1}{|l|}{\#Bugs} & \multicolumn{1}{|l|}{\#Mutants} & \#Bugs & \#Bugs \\ \hline
\multicolumn{1}{|l|}{SpotBugs} & \multicolumn{1}{c|}{868} & \multicolumn{1}{c|}{23,436} & \multicolumn{1}{c|}{10} & \multicolumn{1}{c|}{25,633} & \multicolumn{1}{c|}{12}  & 0 \\ \hline
\multicolumn{1}{|l|}{Infer} & \multicolumn{1}{c|}{339} & \multicolumn{1}{c|}{8,721} & \multicolumn{1}{c|}{2} & \multicolumn{1}{c|}{4,306} & \multicolumn{1}{c|}{2} & 0 \\ \hline

\end{tabular}
}
\end{table}

%% file: sections/threats.tex
\section{Threats to Validity}
\label{sec:threats}

Our approach relies on LLMs to generate seed bugs that can be validated through test cases. However, not all bugs can be effectively triggered via testing. Some bugs may require specific runtime conditions, external dependencies, or particular system configurations that are difficult to replicate in unit tests.  The reliance on test-executable seed bug validation may introduce bias in our findings, as undetectable yet valid bugs might be overlooked. Other strategies that can validate the generated seed bugs should be explored in the future. 
In this work, we examine our approach on three widely adopted static analyzers. However, despite their widespread use in the industry, these analyzers may not represent all static analyzers, limiting the generalizability of our findings. 
Additionally, we focus specifically on validating rules designed to detect bugs in Java programs. As such, our approach may yield different results when applied to other static analyzers or rules targeting bugs in programs written in other programming languages. 
In this work, we experiment with five widely adopted LLMs; however, they may not represent all publicly available LLMs. 
As a result, our approach may yield different results when applied to other LLMs with varying architectures, training data, or reasoning capabilities.

%% file: sections/conclusion.tex
\section{Conclusion}
\label{sec:conclusion}

This paper introduces \textsc{StaAgent}, a novel agentic testing framework that leverages LLMs to automatically generate, mutate, and validate code snippets for testing static analyzers. We evaluated five state-of-the-art LLMs (CodeLlama, DeepSeek, Codestral, Qwen, and GPT-4o) across five widely used static analyzers (SonarQube, SpotBugs, ErrorProne, Infer, and PMD). Our experiments uncovered 64 problematic rules across five analyzers, and 83\% of which cannot be detected by the SOTA baseline approach, demonstrating the effectiveness of LLM-driven data synthesis in identifying implementation flaws and advancing software engineering tools.

In future work, we plan to extend our approach to support static analyzers for additional programming languages and to target other automated software engineering tasks. In addition, we will try other prompting strategies to reduce false-positives and evaluate other models.